\documentclass{sf2a-conf}
\usepackage{graphicx}
\usepackage{bm}

\def\bS{{\bm S}}

\def\lb{\label}

\def\be{\begin{equation}}
\def\ee{\end{equation}}
\def\bea{\begin{eqnarray}}
\def\eea{\end{eqnarray}}

\begin{document}

\TitreGlobal{SF2A 2006}

\title{Testing General Relativity with the ACES Mission}
\author{C. Le Poncin-Lafitte}\address{Lohrmann Observatory, Dresden Technical University, Dresden, Germany.}
\author{S. B. Lambert}\address{Royal Observatory of Belgium, 3 Avenue Circulaire, B-1180 Brussels,
Belgium.}

%
\runningtitle{Testing General Relativity with the ACES Mission}
\setcounter{page}{237}
\index{Teyssandier, p.}
\index{Le Poncin-Lafitte, C.}


\maketitle
\begin{abstract} 
The new generation of atomic clocks will reach unprecedented uncertainties in frequency of $10^{-18}$. In order to prepare space missions such as ACES, we compute all relativistic frequency shifts detectable during this mission in the case of a clock aboard the International Space Station.
\end{abstract}
%
\section{Introduction}

Advances in laser cooling of atoms have led to a new generation of highly accurate atomic clocks (Caesium and Rubidium fountains) having a time-keeping accuracy of the order of $10^{-16}$--$10^{-18}$ in fractional frequency. The main limitating factor to improve the accuracy of ground-based fountains is the Earth gravitational field. That is why it is envisaged to take advantage of microgravity experimental conditions in spacecrafts, especially the International Space Station (ISS). Three space experiments are already planned to be launched in the near future: the NASA's Primary Atomic Reference Clock in Space (PARCS, scheduled for 2008), the Rubidium Atomic Clock Experiment (RACE, expected in early 2008), and the ESA's Atomic Clock Ensemble in Space (ACES, expected for 2010). At this level of accuracy, a complete relativistic framework is indispensable to analyse frequency shift, and it constitutes a great opportunity to test general relativity. Although a large literature is available, no realistic orders of magnitude have been computed in the case of a clock aboard the ISS and passing above a typical mid-latitude observing site. This is exactly what we analyse in this paper.
\section{Relativistic frequency shift}

The frequency shift $\nu_A /\nu_B$ may be written as
\be \lb{66}
\frac{\nu_A}{\nu_B}=1+\left( \frac{\delta \nu}{\nu}\right)_c +
\left( \frac{\delta \nu}{\nu}\right)_g \, ,
\ee
where $(\delta \nu /\nu )_c$ is the special-relativistic Doppler effect and $(\delta \nu/\nu )_g$ contains all the contribution of the gravitational field, eventually mixed with kinetic terms. $(\delta \nu /\nu )_c$ has the following form
\begin{equation}
\left( \frac{\delta \nu}{\nu}\right)_c=\sum_{n=1}^4\frac{1}{c^n}\left( \frac{\delta \nu}{\nu}\right)^{(n)}_c
\end{equation}
In the vicinity of the Earth and for a level of uncertainty around $10^{-18}$ in fractional frequency, we have to consider the Earth as an extended body, but it is sufficient to expand the Earth gravitational field in a multipole. under these assumptions the gravitational frequency shift can be written
\be
\left( \frac{\delta \nu}{\nu}\right)_g=\frac{1}{c^2}\left( \frac{\delta \nu}{\nu}\right)_g^{(2)}+\frac{1}{c^3}\left( \frac{\delta \nu}{\nu}\right)_M^{(3)}+\frac{1}{c^3}\left( \frac{\delta \nu}{\nu}\right)_{J_2}^{(3)}+\frac{1}{c^4}\left( \frac{\delta \nu}{\nu}\right)_M^{(4)}+\frac{1}{c^4}\left( \frac{\delta \nu}{\nu}\right)_\bS^{(4)}+... \, ,
\ee
where $(\delta \nu/\nu)_M^{(3)}$ and $(\delta \nu/\nu)_M^{(4)}$ correspond to the influence of the Earth's mass on Doppler shifht, $(\delta \nu/\nu)_{J_2}^{(3)}$ corresponds to an effect due to the $J_2$, and $(\delta \nu/\nu)_\bS^{(4)}$ is due to the Earth's rotation. The contribution $\left(\delta \nu/\nu\right)_g^{(2)}$ can be written as
\be
\left( \frac{\delta \nu}{\nu}\right)_g^{(2)}=\left( \frac{\delta \nu}{\nu}\right)_{M}^{(2)}+\sum_{n=2}^\infty\left( \frac{\delta \nu}{\nu}\right)_{J_n}^{(2)}\, ,
\ee
where $(\delta \nu/\nu)_M^{(2)}$ correspond to the influence of the Earth's mass and $(\delta \nu/\nu)_{J_n}^{(2)}$ to the contribution of each mass multipole moments on Doppler shifht. Note that analytical expressions are available in litterature (Blanchet \& {\it al} 2001, Linet \& Teyssandier, 2002).

\section{Numerical evaluations}

The frequency shifts incoming from the different terms explicited above are computed numerically in the case of a clock aboard an ISS-like spacecraft and passing above a mid-latitude observing site. For instance, we take the site of Paris, latitude $49^{\circ}$N. The obliquity of the ISS's orbit is around $51^{\circ}$, and its altitude is roughly 350~km, varying by a few tens of kilometers. The Earth is assumed to rotate uniformly around the $z$-axis of a space-fixed frame of reference. Concerning special-relativistic Doppler effects, we compute some maximum values listed below:
\be 
\bigg| \frac{1}{c} \left( \frac{\delta \nu}{\nu}\right)_{c}^{(1)} \bigg| \leq 1\times10^{-5} \, , \quad
\bigg| \frac{1}{c^2} \left( \frac{\delta \nu}{\nu}\right)_{c}^{(2)} \bigg| \leq  2\times 10^{-10} \, ,\quad
\bigg| \frac{1}{c^3} \left( \frac{\delta \nu}{\nu}\right)_{c}^{(3)} \bigg| \leq  3\times 10^{-15} \, ,
\ee 
and
\be 
\bigg| \frac{1}{c^4} \left( \frac{\delta \nu}{\nu}\right)_{c}^{(4)} \bigg| \leq  7\times 10^{-20} \, . 
\ee 
For second order terms $\left( \frac{\delta \nu}{\nu}\right)_g^{(2)}$, we are firstly interested in the effect due to the spherical mass of the Earth. One gets 
\be 
\bigg| \frac{1}{c^2} \left( \frac{\delta \nu}{\nu}\right)_{2}^{(M)} \bigg| \leq  3\times 10^{-11} .
\ee 
Then, we obtain the following upper limit for the influence of the first multipole moments : 
\be 
\bigg| \frac{1}{c^2} \left( \frac{\delta \nu}{\nu}\right)_{J_2}^{(2)} \bigg| \leq  2\times 10^{-13} \, ,\quad 
\bigg| \frac{1}{c^2} \left( \frac{\delta \nu}{\nu}\right)_{J_4}^{(2)} \bigg| \leq  3\times 10^{-16} \, ,\quad 
\bigg| \frac{1}{c^2} \left( \frac{\delta \nu}{\nu}\right)_{J_6}^{(2)} \bigg| \leq  1\times 10^{-16} \, . 
\ee 
Now, we focus on third order terms $\left( \frac{\delta \nu}{\nu}\right)^{(3)}$. We obtain 
\be 
\bigg| \frac{1}{c^3} \left( \frac{\delta \nu}{\nu}\right)_{M}^{(3)} \bigg| \leq  2\times 10^{-14} ,\quad
\bigg| \frac{1}{c^3} \left( \frac{\delta \nu}{\nu}\right)_{J_2}^{(3)} \bigg| \leq  1\times 10^{-18} .
\ee 
Finally, we compute terms of order $O(1/c^4)$ wich are at first sight smallest :
\be 
\bigg| \frac{1}{c^4} \left( \frac{\delta \nu}{\nu}\right)_{M}^{(4)} \bigg| \leq  1\times 10^{-19} \, ,\quad
\bigg| \frac{1}{c^4}\left( \frac{\delta \nu}{\nu}\right)_{\bS}^{(4)} \bigg| 
\leq \, 1\times 10^{-22} \, . 
\ee



\section{Conclusion}
We obtained orders of magnitude of the whole relativistic effects important for ACES. We hope that these evaluations will be useful for the data processing of this mission. Moreover, we can note that relativistic effects are sufficiently numerous to test relativity. However, it should however be noted that they are sensitive to the variability of the ISS orbit which must be controlled perfectly.

\end{document}